\documentclass[iop]{emulateapj}

\usepackage{natbib}
\usepackage{gensymb}
\usepackage{hyperref}
\usepackage{graphicx}
\usepackage{multirow}
\usepackage{array}
\usepackage{makecell}

\newcommand{\mstar}{$M_*$}
\newcommand{\hii}{H~\textsc{ii}\ }
\newcommand{\ii}{~\textsc{ii}}
\newcommand{\iii}{~\textsc{iii}}

\newcommand{\stt}{S$_{32}$}

\shorttitle{The MOSDEF Survey: [S~\textsc{iii}] as a New Probe of Evolving ISM Conditions}
\shortauthors{Sanders et al.}

\begin{document}

\title{The MOSDEF Survey: [S~\textsc{iii}] as a New Probe of Evolving ISM Conditions\altaffilmark{*}}
\altaffiltext{*}{Based on data obtained at the
W.M. Keck Observatory, which is operated as a scientific partnership among the California Institute of Technology, the
University of California, and NASA, and was made possible by the generous financial support of the W.M. Keck Foundation.
}

\author{Ryan L. Sanders\altaffilmark{1}}

\author{Tucker Jones\altaffilmark{1}}

\author{Alice E. Shapley\altaffilmark{2}}

\author{Naveen A. Reddy\altaffilmark{3,4}}

\author{Mariska Kriek\altaffilmark{5}}

\author{Alison L. Coil\altaffilmark{6}}

\author{Brian Siana\altaffilmark{3}}

\author{Bahram Mobasher\altaffilmark{3}}

\author{Irene Shivaei\altaffilmark{7,8}}

\author{Sedona H. Price\altaffilmark{9}}

\author{William R. Freeman\altaffilmark{3}}

\author{Mojegan Azadi\altaffilmark{10}}

\author{Gene C. K. Leung\altaffilmark{6}}

\author{Tara Fetherolf\altaffilmark{3}}

\author{Tom O. Zick\altaffilmark{5}}

\author{Laura de Groot\altaffilmark{11}}

\author{Guillermo Barro\altaffilmark{12}}

\author{Francesca M. Fornasini\altaffilmark{10}}

\altaffiltext{1}{Department of Physics, University of California, Davis, One Shields Ave, Davis, CA 95616, USA}
\altaffiltext{2}{Department of Physics \& Astronomy, University of California, Los Angeles, 430 Portola Plaza, Los Angeles, CA 90095, USA}
\altaffiltext{3}{Department of Physics \& Astronomy, University of California, Riverside, 900 University Avenue, Riverside, CA 92521, USA}
\altaffiltext{4}{Alfred P. Sloan Fellow}
\altaffiltext{5}{Astronomy Department, University of California, Berkeley, CA 94720, USA}
\altaffiltext{6}{Center for Astrophysics and Space Sciences, University of California, San Diego, 9500 Gilman Dr., La Jolla, CA 92093-0424, USA}
\altaffiltext{7}{Department of Astronomy/Steward Observatory, 933 North Cherry Ave, Rm N204, Tucson, AZ, 85721-0065, USA}
\altaffiltext{8}{Hubble Fellow}
\altaffiltext{9}{Max-Planck-Institut f{\"u}r extraterrestrische Physik, Postfach 1312, Garching, 85741, Germany}
\altaffiltext{10}{Harvard-Smithsonian Center for Astrophysics, 60 Garden Street, Cambridge, MA 02138, USA}
\altaffiltext{11}{Department of Physics, The College of Wooster, 1189 Beall Avenue, Wooster, OH 44691, USA}
\altaffiltext{12}{Department of Phyics, University of the Pacific, 3601 Pacific Ave, Stockton, CA 95211, USA}

\email{email: rlsand@ucdavis.edu}

\begin{abstract}
We present measurements of [S\iii]$\lambda\lambda$9069,9531 for a sample of $z\sim1.5$ star-forming galaxies,
 the first representative sample with measurements of these lines at $z\gtrsim0.1$.
We employ the line ratio \stt$\equiv$[S\iii]$\lambda\lambda$9069,9531/[S\ii]$\lambda\lambda$6716,6731 as a novel probe of evolving ISM conditions.
Since this ratio includes the low-ionization line [S\ii], it is crucial that the effects of diffuse ionized gas (DIG)
 on emission-line ratios be accounted for in $z\sim0$ galaxy spectra, or else that comparisons be made to samples of
 local \hii regions in which DIG emission is not present.
We find that \stt\ decreases with increasing stellar mass at both $z\sim1.5$ and $z\sim0$, but with a shallow slope
 suggesting \stt\ has a weak dependence on metallicity,
 in contrast with [O\iii]/[O\ii] that displays a strong metallicity dependence.
As a result, \stt\ only mildly evolves with redshift at fixed stellar mass.
The $z\sim1.5$ sample is systematically offset towards lower \stt\ and higher [S\ii]/H$\alpha$ at fixed [O\iii]/H$\beta$ relative
 to $z=0$ \hii regions.
We find that such trends can be explained by
 a scenario in which the ionizing spectrum is harder at fixed O/H with increasing redshift,
 but are inconsistent with an increase in ionization parameter at fixed O/H.
This analysis demonstrates the advantages of expanding beyond the strongest rest-optical lines for evolutionary studies,
 and the particular utility of [S\iii] for characterizing evolving ISM conditions and stellar compositions.
These measurements provide a basis for estimating [S\iii] line strengths for high-redshift galaxies,
 a line that the {\it James Webb Space Telescope} will measure out to $z\sim5.5$.
\end{abstract}


\section{Introduction}\label{sec:intro}

The rest-optical emission lines of star-forming galaxies provide valuable insight into the physical properties
 of the ionized gas in \hii regions.
Diagnostics of nebular metallicity (O/H), ionization parameter ($U$),
 and electron density have been calibrated at $z\sim0$.
An additional important consideration is the shape of the stellar spectrum ionizing the gas, primarily determined
 by the Fe/H of massive stars.
Due to tight relations between these properties, \hii regions and local
 star-forming galaxies follow excitation sequences in emission-line ratio diagrams, including the
 [O\iii]/H$\beta$ vs.\ [N\ii]/H$\alpha$ and [S\ii]/H$\alpha$ ``BPT" diagrams \citep{bpt81,vei87}.
Recent large near-infrared spectroscopic surveys at $z\sim2$ have demonstrated that high-redshift star-forming galaxies
 follow a sequence that is systematically offset from the $z\sim0$ sequence in the [N\ii] BPT diagram
 \citep[e.g.,][]{ste14,kas19a,san16a,sha19}.
This shift in the excitation sequence implies that the ionized interstellar medium (ISM) conditions are changing with
 redshift, and in particular that at least some of the relations between O/H, $U$, and stellar Fe/H evolve.

Determining which parameters evolve at fixed O/H and by how much has proven difficult.
Such work often relies on photoionization models to understand how each line ratio changes when
 the relevant ISM conditions are varied \citep[e.g.,][]{kew13,ste14,san16a,str17}.
A particular challenge lies in characterizing the ionization parameter and hardness of the ionizing
 spectrum as neither is simply tied to an observable and the two are highly degenerate when
 O/H is unknown.
Useful constraints cannot be obtained when only a few rest-optical lines are available (the case for
 most high-redshift galaxies), and degeneracies and large uncertainties remain even when 5$-$6 strong
 rest-optical lines are available \citep{str18}.
Only one ``pure" ionization parameter diagnostic is available among the strongest optical lines ([O\iii]/[O\ii])
 and it is strongly affected by dust reddening.
A better understanding of evolving ISM conditions can be obtained by introducing additional line ratios as
 constraints that move beyond the strongest optical lines ([O\ii], H$\beta$, [O\iii], H$\alpha$, [N\ii], [S\ii]),
 most preferrably an additional ionization parameter diagnostic ratio of a single element that is relatively
 unaffected by reddening. 

In this letter, we present the first measurements of [S\iii]$\lambda\lambda$9069,9531
 for a representative sample at $z>1$ and explore the utility of the \stt$\equiv$[S\iii]$\lambda\lambda$9069,9531/[S\ii]$\lambda\lambda$6716,6731
 ratio for constraining evolving ISM conditions.
This analysis uses observations of star-forming galaxies at $z\sim1.5$ from the
 MOSFIRE Deep Evolution Field survey \citep[MOSDEF;][]{kri15} in combination with galaxy
 and \hii region samples from the local universe.
In \S\ref{sec:data}, we present the [S\iii] detections for individual high-redshift galaxies and composite spectra, and
 describe the low-redshift comparison samples.
We investigate the evolution of \stt\ at fixed stellar mass (\mstar) in \S\ref{sec:mstar}.
Finally, we present excitation sequences of [S\iii] and [S\ii] ratios in \S\ref{sec:ism} and discuss implications
 for the evolving physical conditions in \hii regions.
Emission-line wavelengths are given in air.
We assume a $\Lambda$CDM cosmology with H$_0$=70~km~s$^{-1}$~Mpc$^{-1}$, $\Omega_{\text{m}}$=0.3, and
 $\Omega_{\Lambda}$=0.7.

\section{Observations, Data, \& Samples}\label{sec:data}

\subsection{The $z\sim1.5$ MOSDEF sample}

We draw a sample of high-redshift galaxies with [S\iii] measurements from the MOSDEF survey,
 a four-year program that obtained rest-frame optical spectra of $\sim1,500$ galaxies at $1.37\le z\le3.80$.
A detailed description of the survey and data reduction can be found in \citet{kri15}.
The MOSDEF flux calibration is robust between filters,
 such that ratios of lines from different filters (e.g., [S\iii]/[S\ii]) are biased less than
 0.05~dex on average with a scatter of 0.07~dex \citep{kri15}.
We utilize emission-line measurements, stellar masses, reddening estimates, and star-formation rates (SFRs)
 from the MOSDEF catalogs.
Stellar masses and continuum reddening are estimated from emission-line corrected photometry
 via SED fitting using the code FAST \citep{kri09}, assuming constant star-formation histories, solar metallicity,
 a \citet{cha03} initial mass function (IMF), and the \citet{cal00} attenuation curve.
Nebular reddening is estimated using the H$\alpha$/H$\beta$ ratio when both lines are detected at S/N$\ge$3,
 assuming an intrinsic ratio of 2.86 and the \citet{car89} extinction curve.
When H$\beta$ is not detected, we infer E(B-V)$_{\text{gas}}$ from continuum reddening under the assumption that
 E(B-V)$_{\text{gas}}$$\approx$E(B-V)$_{\text{stars}}$, as found to be true on average at $z\sim2$ \citep{kas13,red15}.
SFRs are derived from reddening-corrected H$\alpha$ luminosity using the \citet{hao11} calibration converted to
 a \citet{cha03} IMF.

\begin{figure}
 \centering
 \includegraphics[width=\columnwidth]{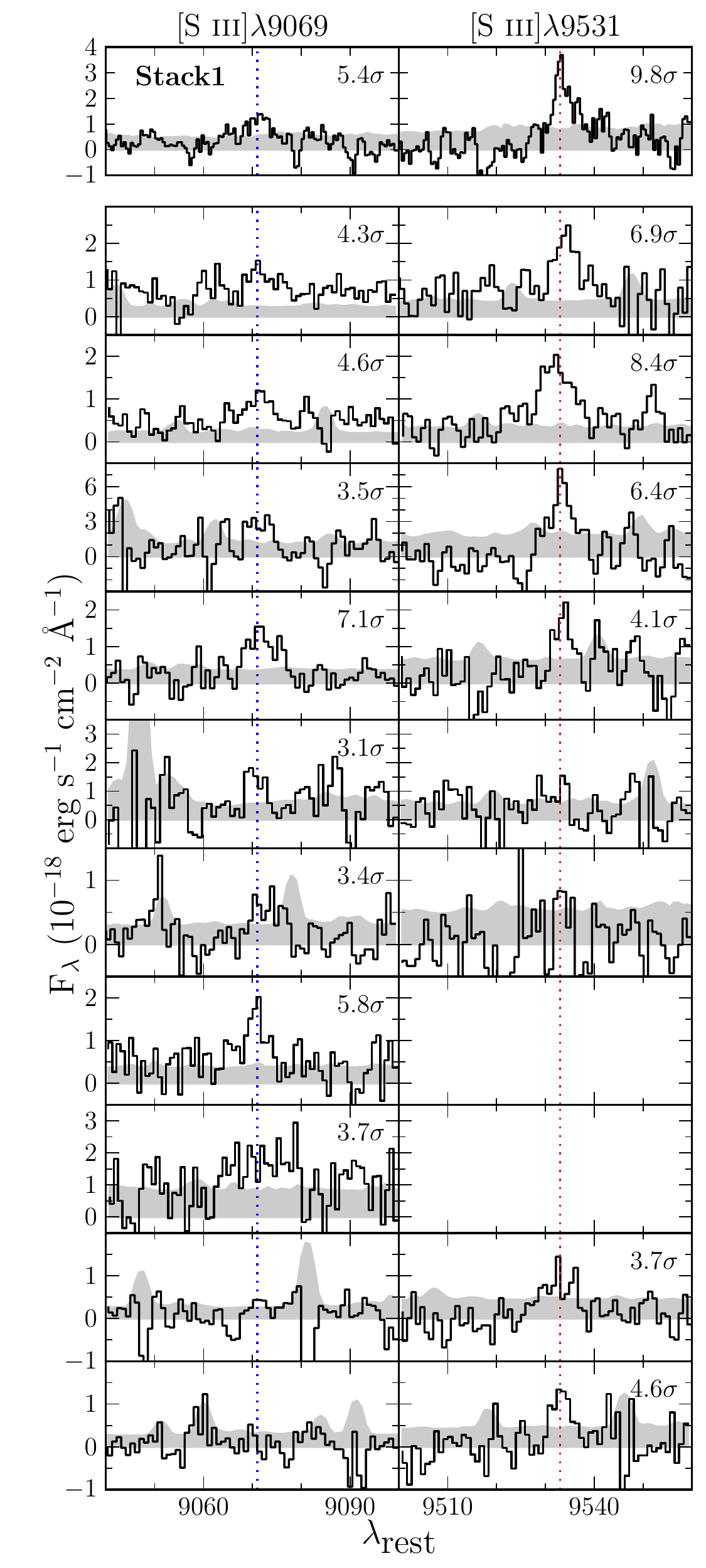}
 \caption{
Science spectra (black) and error spectra (gray) displaying detections of [S\iii]$\lambda$9069 (left column)
 and [S\iii]$\lambda$9531 (right column) at $z\sim1.5$ from the MOSDEF survey.
All rows except the top show spectra of individual star-forming galaxies.
The top row presents the {\it Stack1} composite spectrum with flux arbitrarily normalized.
The significance of the emission-line detection is given in the top right corner of panels when
 S/N$\ge$3.
}\label{fig1}
\end{figure}

Galaxies were targeted in three redshift bins: $1.37\le z\le1.70$, $2.09\le z\le2.61$,and $2.95\le z\le3.80$.
Masks in each redshift bin were observed in multiple near-infrared filters in which the [O\ii], H$\beta$, [O\iii],
 H$\alpha$, [N\ii], and [S\ii] emission lines fall.
Accordingly, $z\sim1.5$ masks were observed in $Y$, $J$, and $H$ bands; $z\sim2.3$ masks in $J$, $H$, and $K$;
 and $z\sim3.4$ masks in $H$ and $K$ only.
[S\iii] lies significantly redwards of [S\ii], falling in $K$ band at $z\sim1.5$ and redshifted out of the bands of atmospheric
 transmission at $z>2$.
Thus, most MOSDEF $z\sim1.5$ targets do not have [S\iii] measurements because observations in $K$ band are lacking.
However, there are 49 MOSDEF star-forming galaxies at $1.25\le z\le1.65$ with $K$-band observations because they were either filler
 targets (26) or serendipitous detections (23) on $z>2$ masks.
At least one [S\iii] line is detected  at S/N$\ge$3 for 10 individual galaxies in this sample, with both lines detected in four.
The spectra of these 10 objects are presented in Figure~\ref{fig1}.
In addition to individual detections and limits, we stacked spectra to obtain average [S\iii] measurements for the sample and
 include non-detections.
We created multiple composite spectra following the procedure of \citet{san18}, requiring H$\alpha$ S/N$\ge$3 as well
 as the additional criteria described in Table~\ref{tab:stacks} for each.
In brief, individual spectra were converted to luminosity density, normalized by H$\alpha$ luminosity,
 shifted onto a common rest-wavelength grid, and median stacked without weighting.
This process ensures that high-SFR galaxies do not dominate the composite line ratios.
The {\it Stack1} composite spectrum with coverage of both [S\iii] lines is displayed in the top row of Figure~\ref{fig1}.

\renewcommand{\cellalign}{cl}

\begin{table}
 \centering
 \caption{Description of the $z\sim1.5$ [S\iii] composite spectra.
 }\label{tab:stacks}
 \begin{tabular}{ l l l l }
   \hline\hline
   Name & Selection Criteria & N$_{\text{gal}}$$^a$ & $\langle \log{(\frac{M_*}{\mbox{M}_\odot})}\rangle$$^b$ \\
   \hline\hline
   {\it Stack1} & \makecell{[S\iii]$\lambda$9069 and \\ {[}S\iii]$\lambda$9531 coverage} & 28 & 9.78 \\
   \hline
   {\it Stack2} & \makecell{[S\iii]$\lambda$9069 coverage \\ and 2 bins in \mstar} & 35, 11$^{c}$ & 9.54, 10.21$^{c}$ \\
   \hline
   {\it Stack3} & \makecell{[S\iii]$\lambda$9069, [O\iii]$\lambda$5007, \\ and H$\beta$ coverage} & 29 & 9.85 \\
   \hline
 \end{tabular}
 \begin{flushleft}
 $^{a}$ {The number of galaxies included in each composite.} \newline
 $^{b}$ {The median stellar mass of the galaxies in each composite.} \newline
 $^{c}$ {The low-mass bin is listed first, followed by the high-mass bin. The two mass bins, split at
 10$^{9.94}$~M$_{\odot}$, were populated such that equivalent S/N is obtained on [S\iii]$\lambda$9069 in both composite spectra.}
 \end{flushleft}
\end{table}

\subsection{The $z\sim0$ comparison samples}

Most large spectroscopic $z\sim0$ galaxy surveys \citep[e.g., SDSS;][]{yor00} do not have coverage extending
 to $\approx$1~$\mu$m, required for [S\iii] measurements.
To obtain a $z\sim0$ galaxy comparison sample, we use data from the MaNGA integral field spectroscopic survey
 \citep{bun15}, with coverage out to 1.04~$\mu$m.
We employ the MaNGA PIPE3D catalog of emission-line measurements \citep{sanc16,sanc18}, which includes
 spatially-resolved line fluxes and uncertainties as well as tabulated [O\iii]/H$\beta$ and [N\ii]/H$\alpha$ ratios
 in the central 2.5\arcsec\ of each galaxy.
We select star-forming galaxies using the central line ratios based on the demarcation of \citet{kau03}
 and restrict the redshift to $z<0.08$ to ensure that [S\iii]$\lambda$9531 falls in the bandpass, yielding
 a sample of 1,150 star-forming galaxies with a median redshift of $z_{\text{med}}=0.026$.
The MaNGA sample has a lower median redshift than typical SDSS star-forming galaxy samples with $z_{\text{med}}\approx0.07-0.10$
 \citep[e.g.,][]{tre04,and13}.
To obtain integrated galaxy spectra similar to SDSS fiber spectra, we sum the MaNGA line fluxes within the central
 10\arcsec$\times$10\arcsec\ corresponding to a 5~kpc width at $z=0.026$, equivalent to the physical diameter covered by
 a 3\arcsec\ SDSS fiber at $z=0.085$.

We also compare to $z=0$ \hii regions in three spiral galaxies from the CHAOS survey
 (NGC~628, \citealt{ber15}; NGC~5194, \citealt{cro15}; NGC~5457, \citealt{cro16}).
This sample comprises 213 individual \hii regions with detections of both [S\iii] lines,
 spanning a wide range of metallicity ($0.1-2.0$~Z$_{\odot}$).

\subsection{The [S\iii]$\lambda$9531/$\lambda$9069 ratio}

\begin{figure}
 \includegraphics[width=\columnwidth]{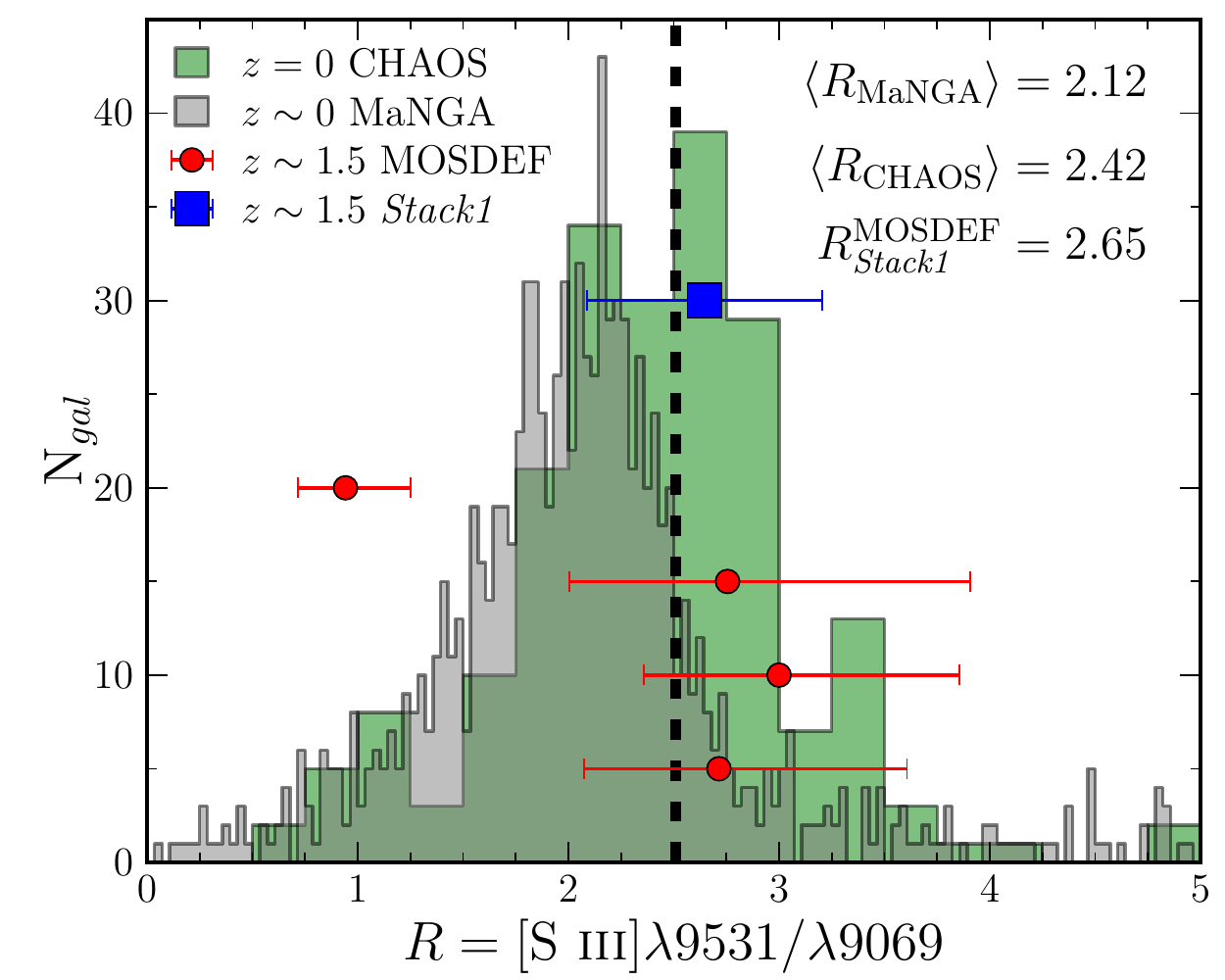}
 \centering
 \caption{
Histogram of [S\iii]$\lambda$9531/$\lambda$9069 for $z=0$ \hii regions from the CHAOS survey (green)
 and $z\sim0$ star-forming galaxies from the MaNGA survey (gray).
Red points show the four $z\sim1.5$ MOSDEF star-forming galaxies with S/N$\ge$3 detections of both lines,
 placed at arbitrary y-axis positions.
The blue square presents the ratio measured from the {\it Stack1} composite spectrum.
The {\it Stack1} ratio and median values for the $z\sim0$ samples are given in the top right corner.
The dashed vertical line shows the theoretically predicted value of 2.5.
}\label{fig2}
\end{figure}

Not all of our targets are detected in both [S\iii] lines.
Accordingly, we need to convert the line flux of one line to the total doublet flux.
This approach is possible because the ratio [S\iii]$\lambda$9531/$\lambda$9069 is fixed to a value of $\approx$2.5
 according to the transition probabilities \citep{ost06,tay19}.
We show the [S\iii]$\lambda$9531/$\lambda$9069 ratios for individual $z\sim1.5$ galaxies, the {\it Stack1}
 $\sim1.5$ composite, and the $z\sim0$ comparison samples in Figure~\ref{fig2}.
We find that {\it Stack1} and three out of four $z\sim1.5$ galaxies are consistent with the theoretically expected
 ratio within 1$\sigma$.  The single offset $z\sim1.5$ galaxy displays possible skyline contamination on the red wing of
 [S\iii]$\lambda$9531 (Fig.~\ref{fig1}, fifth row).
The $z=0$ CHAOS \hii regions also match the expected value on average.
The median ratio of the $z\sim0$ MaNGA sample is 2.12, significantly lower than the expected value.
This offset may indicate
 that Paschen-$\epsilon$ absorption at 9546~\AA\ has not been fully accounted for.
To avoid biasing the total [S\iii] fluxes low, we only use [S\iii]$\lambda$9069 for the MaNGA sample.
For all samples and composites, we infer the total [S\iii]$\lambda\lambda$9069,9531 flux assuming that
 [S\iii]$\lambda$9531/$\lambda$9069=2.5 when only one [S\iii] line is detected.

\section{S$_{32}$ and global galaxy properties}\label{sec:mstar}

\begin{figure}
 \includegraphics[width=\columnwidth]{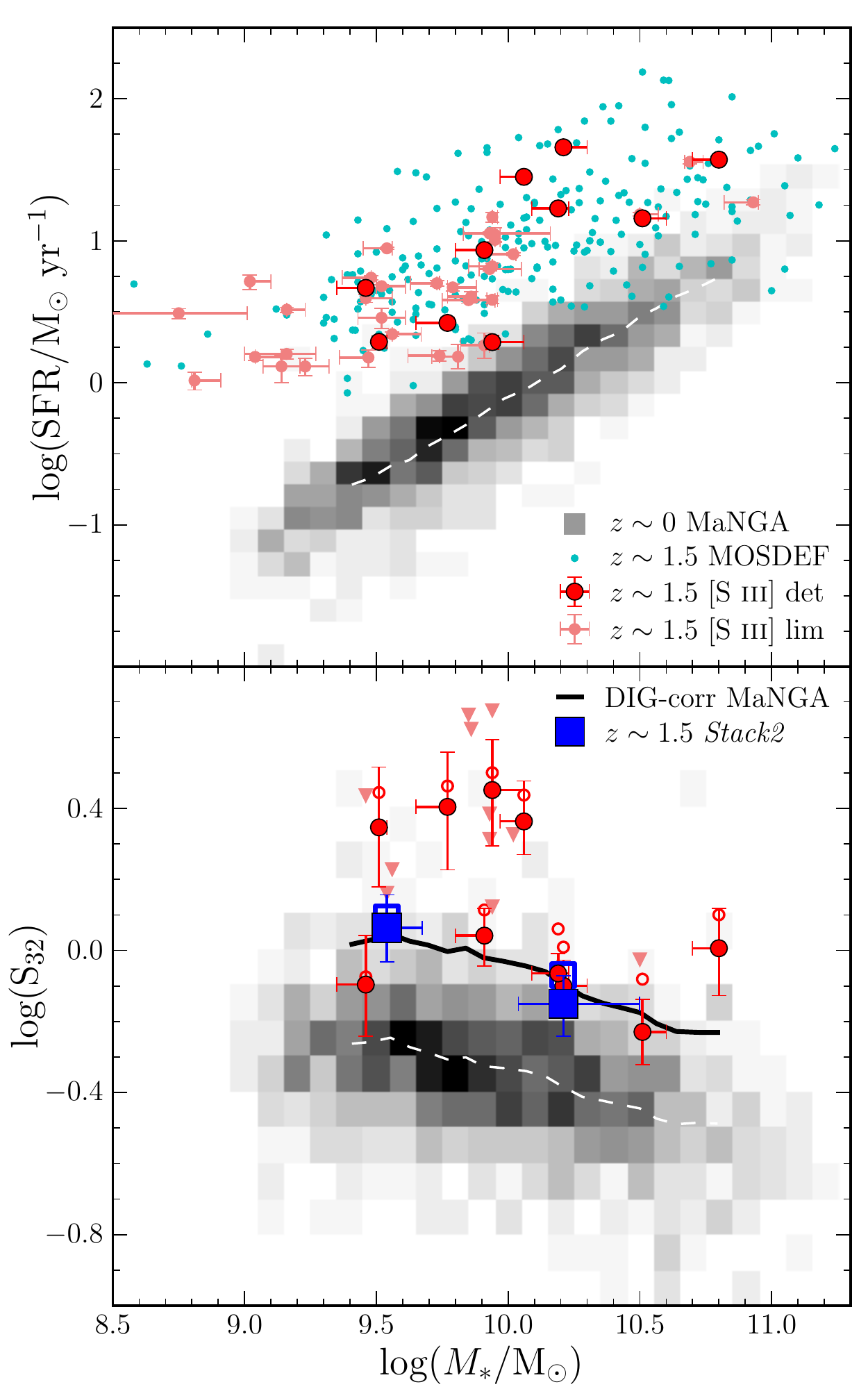}
 \centering
 \caption{
Star-formation rate (top) and \stt$\equiv$[S\iii]/[S\ii] (bottom) vs.~stellar mass.
Filled red points show individual $z\sim1.5$ star-forming galaxies with detections of at least one [S\iii] line.
In the top panel, pink circles denote $z\sim1.5$ galaxies with wavelength coverage of [S\iii]
 but without detections, while pink triangles show upper limits on \stt\ in the lower panel.
Cyan points in the top panel present the full $z\sim1.5$ MOSDEF star-forming sample with H$\alpha$ S/N$\ge$3.
The gray two-dimensional histogram shows the distribution of the $z\sim0$ MaNGA sample, with
 running medians displayed as white dashed lines.
The black solid line in the bottom panel shows the MaNGA median after correcting [S\ii] emission for DIG contamination.
Blue filled squares in the bottom panel present the low- and high-mass composites of {\it Stack2}.
Open blue and red points denote \stt\ values prior to reddening correction.
}\label{fig3}
\end{figure}

We now investigate the evolution of the emission-line ratio
 \stt$\equiv$[S\iii]$\lambda\lambda$9069,9531/[S\ii]$\lambda\lambda$6716,6731.
We begin by characterizing the global galaxy properties of the $z\sim1.5$\ [S\iii] sample.
In the top panel of Figure~\ref{fig3}, we show SFR vs.\ \mstar\ for the $z\sim1.5$ MOSDEF and
 $z\sim0$ MaNGA samples.
The $z\sim1.5$ galaxies with [S\iii] detections (red) and those with [S\iii] coverage but no
 detections (pink) fall on the mean $z\sim1.5$ \mstar-SFR relation described by the full
 MOSDEF $z\sim1.5$ star-forming sample (cyan).
The [S\iii] subset has a lower average \mstar\ ($\sim$10$^{9.7}$~M$_{\odot}$) than the full
 MOSDEF sample ($\sim$10$^{10.0}$~M$_{\odot}$), but is not significantly biased in SFR at fixed \mstar.
Both individual [S\iii]-detected galaxies and stacks including non-detections are representative
 of the typical $z\sim1.5$ star-forming population.

The bottom panel of Figure~\ref{fig3} displays \stt\ vs.\ \mstar.
Individual $z\sim1.5$ galaxies span a wide range of \stt, with the highest \stt\ values occurring in the low-mass
 half of the sample.
Blue squares show stacks of $z\sim1.5$ spectra in two \mstar\ bins ({\it Stack2}).
At fixed \mstar, [S\iii]-detected galaxies lie at higher \stt\ than the stacks
 due to preferential detection of high-excitation galaxies with stronger [S\iii].
We find that \stt\ decreases with increasing \mstar\ for both the $z\sim1.5$ and $z\sim0$ samples, indicative of
 lower ionization parameter at higher \mstar\ and higher metallicity.
The anticorrelation between \stt\ and \mstar\ displays a similar slope at $z\sim1.5$ and $z\sim0$,
 but the $z\sim1.5$ galaxies are offset 0.25~dex higher in \stt\ at fixed \mstar.

It has been shown that [S\ii] is significantly enhanced in $z\sim0$ integrated galaxy spectra due to diffuse ionized gas (DIG)
 emission, which is expected to be negligible at high redshifts \citep{zha17,san17,sha19}.
DIG contamination thus biases redshift evolution comparisons.
We correct for DIG contamination of [S\ii] in the MaNGA line ratios according to the prescription of \citet{san17}, removing
 the DIG contribution and yielding the contribution from \hii regions only.
The effect of DIG on galaxy-integrated [S\iii] emission has not yet been investigated.
However, DIG contribution to [S\iii] is not expected to be strong since DIG primarily enhances low-ionization species.
After correcting the MaNGA sample for DIG contamination of [S\ii], the relation between \stt\ and \mstar\ appears to be nearly the same
 at $z\sim0$ (black line) and $z\sim1.5$ (blue squares).

Many strong-line ratios (e.g., [O\iii]/H$\beta$, [O\iii]/[O\ii]) display significant
 evolution towards higher excitation at fixed \mstar\ even after accounting for $z\sim0$ DIG \citep{san16a,san18},
 thought to primarily reflect evolution towards lower metallicity at fixed \mstar\ with increasing redshift.
We now examine whether the observed lack of significant \stt\ evolution after DIG correction is consistent with this picture.
Mass-metallicity relation studies at both $z\sim0$ and $z\sim1.5$ find $\Delta$log(O/H)$\sim$0.6~dex over a decade in \mstar\ at fixed redshift \citep{zah14b,kas17}, while Figure~\ref{fig3} shows \stt\ changes by $\sim0.2$~dex over the same mass interval.
Accordingly, \stt\ has a very weak dependence on metallicity ($\Delta\log(\mbox{S}_{32})\approx0.33\times\Delta\log(\mbox{O/H})$).
Indeed, \stt\ displays a much weaker dependence on metallicity than [O\iii]/[O\ii] in photoionization models
 \citep{kew19}.
From $z\sim0$ to $z\sim1.5$, O/H decreases by $\sim$0.2~dex at fixed \mstar\ corresponding to an expected increase
 in \stt\ of only $\sim0.07$~dex.
This difference is smaller than the {\it Stack2} error bars.
We thus find that the lack of significant \stt\ evolution at fixed \mstar\ after DIG correction is fully
 consistent with the evolution of the mass-metallicity relation.

\section{S$_{32}$ and evolving ISM conditions}\label{sec:ism}

\begin{figure*}
 \includegraphics[width=0.8\textwidth]{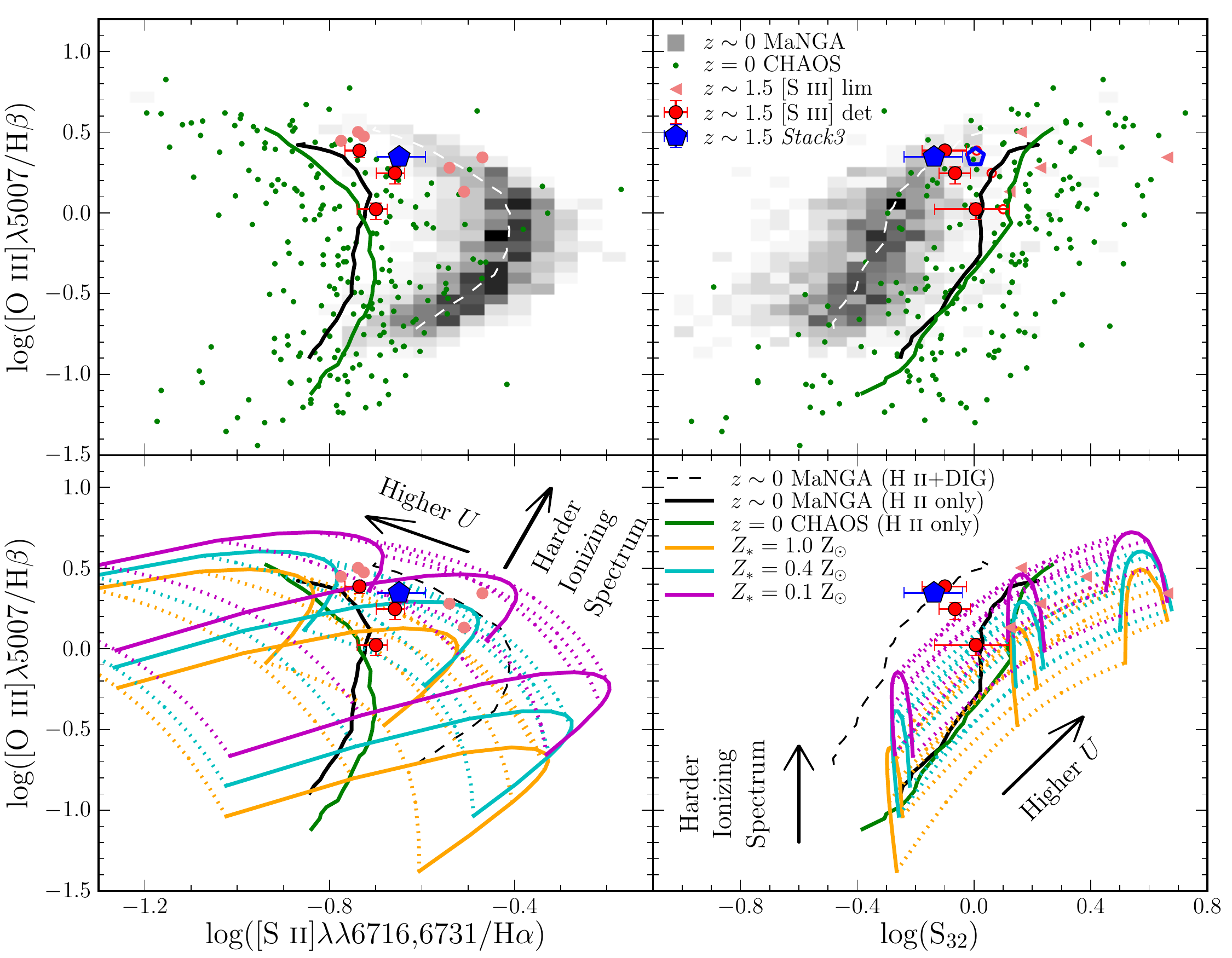}
 \centering
 \caption{
Emission-line ratio diagrams of [O\iii]$\lambda$5007/H$\beta$ vs.\ [S\ii]$\lambda\lambda$6716,6731/H$\alpha$
 (left column) and \stt$\equiv$[S\iii]/[S\ii] (right column) with observations in the top panels and theoretical photoionization
 model grids in the bottom panels.
In the top panels, green points show individual $z=0$ \hii regions from the CHAOS survey.
The blue filled pentagon presents the line ratios of the $z\sim1.5$ {\it Stack3} composite,
 while the open pentagon shows the values without reddening correction.
All other points are as in Figure~\ref{fig3}.
The green and white dashed lines display the running median as a function of [O\iii]$\lambda$5007/H$\beta$ for the CHAOS
 and MaNGA samples, respectively.
The solid black line shows the MaNGA median after correcting [S\ii] and [O\iii] emission for DIG contamination.
In the bottom panels, photoionization models grids are displayed, color-coded by stellar metallicity that is
 a proxy for the hardness of the ionizing spectrum.
Lines of constant ionization parameter are solid, while lines of constant metallicity are dotted.
Black arrows show the qualitative shift when varying only ionization parameter or the hardness of the ionizing spectrum.
The $z\sim0$ medians and $z\sim1.5$ points from the top panels are included for comparison to the model grids, where the
 $z\sim0$ MaNGA median is now dashed black for visibility.
The offset between the $z\sim1.5$ sample and the $z=0$ \hii regions or DIG-corrected MaNGA sample can be reproduced with
 a harder ionizing spectrum at fixed O/H.
}\label{fig4}
\end{figure*}

We now turn to the evolution of excitation sequences in emission-line ratio diagrams and the changing
 the physical conditions of ionized gas in \hii regions with redshift.
In Figure~\ref{fig4}, we show [O\iii]/H$\beta$ vs.\ [S\ii]/H$\alpha$ (left column) and \stt\ (right column).
The top row displays empirical data sets.
In the top left panel, the $z\sim1.5$ sample displays larger [S\ii]/H$\alpha$ at fixed [O\iii]/H$\beta$ than
 $z=0$ \hii regions on average, but is offset towards lower [S\ii]/H$\alpha$ compared to the $z\sim0$ MaNGA galaxies.
The severe offset between the \hii regions and $z\sim0$ galaxies demonstrates the strong influence of DIG
 emission on [S\ii] \citep{san17,sha19}.
Correcting [S\ii] and [O\iii] emission for DIG contamination following \citet{san17}
 yields the black line that closely matches the median sequence of the \hii regions.
In the top right panel, the $z\sim1.5$ sample lies on the sequence described by $z\sim0$ MaNGA galaxies, but is
 offset from the $z=0$ \hii region sequence towards lower \stt\ at fixed [O\iii]/H$\beta$.
Once again, the influence of DIG emission biases the comparison to low-redshift galaxies.

To interpret these offsets, we employ the set of photoionization models described in \citet{san19}
 to identify the qualitative shift in these line ratios when varying \hii region physical conditions.
These models were run using Cloudy \citep{fer17} with BPASS v2.2.1 binary models \citep{sta18}
 as the input radiation field,
 where the stellar metallicity ($Z_*$=Fe/H) is a proxy for the hardness of the ionizing spectrum.
The grids span $0.05-1.5$~Z$_{\odot}$ in nebular metallicity and log$U$=$-2.5$ to $-3.5$,
 and are color-coded by $Z_*$.
Lines of constant ionization parameter are solid and lines of constant O/H are dotted.
The models adopt an electron density of $n_e=250$~cm$^{-3}$, typical at $z\sim2$ \citep{san16a}.
Varying $n_e$ between 25~cm$^{-3}$ (typical at $z\sim0$) and 250~cm$^{-3}$ changes
 [O\iii]/H$\beta$ and [S\ii]/H$\alpha$ by $\lesssim$0.05~dex and does not change \stt,
 and does not affect our results.

The grids fail to fully overlap the $z\sim1.5$ sample in the bottom right panel
 of Figure~\ref{fig4}.
This failure is likely due to a known problem wherein photoionization models underproduce [S\ii] relative
 to other lines \citep[see][and references therein]{kew19}, preventing us from
 placing quantitative constraints on the gas properties.
However, the qualitative direction of line ratio shifts should be robust even in the face of systematic
 [S\ii] underestimation.
We thus proceed by comparing the observed offsets between the $z\sim0$ and $z\sim1.5$ samples in Figure~\ref{fig4}
 with the qualitative shifts in model line ratios when varying ionization parameter and the hardness of the
 ionizing spectrum.

We first consider the scenario where the ionizing spectrum varies with redshift at fixed O/H
 (i.e., moving between colors at a fixed grid vertex), 
We find that a harder ionizing spectrum
 (lower $Z_*$) leads to higher [O\iii]/H$\beta$ and [S\ii]/H$\alpha$ while leaving \stt\ unchanged.
As a result, excitation sequences shift towards higher [S\ii]/H$\alpha$ and lower
 \stt\ at fixed [O\iii]/H$\beta$, in agreement with the observed offsets between $z\sim1.5$ galaxies
 and $z=0$ \hii regions or DIG-corrected $z\sim0$ MaNGA galaxies.
We conclude that the shift in line-ratio excitation sequences between $z\sim0$ and $z\sim1.5$ is primarily
 driven by a harder ionizing spectrum at fixed nebular metallicity and does not require significant changes
 to $U$ at fixed O/H.

If we instead consider varying the ionization parameter
 while keeping all other parameters fixed, increasing $U$ (i.e., moving along dotted lines of a single color)
 leads to an increase in [O\iii]/H$\beta$ and \stt, and a decrease in [S\ii]/H$\alpha$.
The net effect is to shift galaxies towards lower [S\ii]/H$\alpha$ at fixed [O\iii]/H$\beta$, and
 along the [O\iii]/H$\beta$$-$\stt\ sequence producing no significant offset in \stt\ at fixed [O\iii]/H$\beta$
 since lines of constant metallicity
 run roughly parallel to the full empirical sequences in the lower right panel.
Thus, larger $U$ at fixed O/H can account for the offset (or lack thereof) between the $z\sim1.5$ sample and the $z\sim0$
 MaNGA sample {\it without DIG correction} (dashed black lines in the lower panels).
This conclusion was reached by past studies based on the position of $z\sim1.6$ star-forming galaxies in the [SII] BPT
 diagram relative to a $z\sim0$ SDSS sample in which DIG was not accounted for \citep{kas17,kas19a}. 
However, since DIG emission is expected to be neglibible in the highly star-forming compact galaxies at high redshifts \citep{sha19},
 a fair comparison is either to the \hii region sample or to DIG-corrected integrated galaxy spectra, from which the
 $z\sim1.5$ galaxies are offset towards {\it higher} [S\ii]/H$\alpha$ and {\it lower} \stt\ at fixed [O\iii]/H$\beta$.
Higher $U$ at fixed O/H fails to account for these offsets.

Our results thus favor a harder ionizing spectrum at fixed O/H with increasing redshift,
 in agreement with recent work at $z\sim2$ based on deep rest-UV continuum spectroscopy \citep{ste16} and electron
 temperature metallicities \citep{san19}.
However, the measurements utilized in this work can be acquired for many galaxies with
 a significantly smaller observational investment.
Increasing the sample of $z>1$ galaxies with [S\iii] detections thus presents a viable path forward to an understanding
 of ISM conditions in individual high-redshift galaxies spanning a wide range in \mstar, SFR, and metallicity.
Fully leveraging new [S\iii] observations requires more realistic photoionization models
 to turn qualitative conclusions into quantitative constraints.
Photoionization modeling will be particularly discriminating in cases where both \stt\ and [O\iii]/[O\ii] are available,
 where the use of two independent ionization parameter diagnostics simultaneously can break degeneracies between the
 ionizing spectrum and $U$.
In the next few years, obtaining high-redshift data sets that expand beyond the strongest rest-optical emission lines
 should be a priority.
Such observations are necessary to develop the tools required to interpret the wealth of information
 from the NIRSpec instrument on the {\it James Webb Space Telescope},
 which will have sufficient sensitivity and wavelength coverage to measure [S\iii] and numerous
 other weak lines up to $z\sim5.5$.

\acknowledgements 
We acknowledge support from NSF AAG grants AST-1312780, 1312547, 1312764, and 1313171,
 grants AR-13907 and GO-15077 provided by NASA through the Space Telescope Science Institute,
 and grant NNX16AF54G from the NASA ADAP program.
We also acknowledge a NASA
contract supporting the “WFIRST Extragalactic Potential
Observations (EXPO) Science Investigation Team” (15-WFIRST15-0004), administered by GSFC.
We wish to extend special thanks to those of Hawaiian ancestry on
 whose sacred mountain we are privileged to be guests. Without their generous hospitality,
 the work presented herein would not have been possible.


\end{document}